\documentclass[fleqn,10pt]{PARCTwoColumn}

\usepackage{listings}
\usepackage{color}
\usepackage{textcomp}
\usepackage{xspace}
\usepackage{url}
\usepackage{gb4e}
\usepackage{framed}
\usepackage{algpseudocode}
\usepackage{hyphenat}

\newcommand{\daemon}{\texttt{metis\_daemon}\xspace}
\newcommand{\control}{\texttt{metis\_control}\xspace}
\newcommand{\ioops}{\texttt{MetisIoOps}\xspace}

\newcommand{\hyperlabel}[1]{}
\newcommand{\nolinkurl}[1]{}
\newcommand{\breakitem}[1]{{\parbox[t]{1.1\linewidth}{\hangindent=13pt {#1}}}}

\newcommand{\fname}[1]{\emph{#1}}

\Masthead{Networking and Distributed Systems\\Computing Science Laboratory}

\PaperTitle{Metis CCNx 1.0 Forwarder} 

\Authors{Marc Mosko*\textsuperscript{1}} 
\affiliation{\textsuperscript{1}\textit{Computing Science Laboratory, PARC}} 
\affiliation{*\textbf{Corresponding author}: marc.mosko@parc.com} 

\newcommand{\keywordname}{Keywords} 

\Abstract{Metis is the CCNx 1.0 forwarder that implements the CCNx 1.0 Semantics and Messages
draft standards.  This document describes how to use Metis and the internal software architecture.}

\begin{document}

	
\definecolor{listinggray}{gray}{0.9}
\definecolor{lbcolor}{rgb}{0.9,0.9,0.9}
\lstset{
	linewidth=\linewidth,
	xleftmargin=20pt,
	resetmargins=true,
	backgroundcolor=\color{lbcolor},
	tabsize=4,    
	language=C,
        basicstyle=\scriptsize,
        upquote=true,
        aboveskip={1.5\baselineskip},
        columns=fixed,
        showstringspaces=false,
        extendedchars=false,
        breaklines=true,
        prebreak = \raisebox{0ex}[0ex][0ex]{\ensuremath{\hookleftarrow}},
        frame=single,
        numbers=left,
        showtabs=false,
        showspaces=false,
        showstringspaces=false,
        identifierstyle=\ttfamily,
        keywordstyle=\color[rgb]{0,0,1},
        commentstyle=\color[rgb]{0.026,0.112,0.095},
        stringstyle=\color[rgb]{0.627,0.126,0.941},
        numberstyle=\color[rgb]{0.205, 0.142, 0.73},
}

\lstset{
	linewidth=\linewidth,
	xleftmargin=20pt,
	resetmargins=true,
	backgroundcolor=\color{lbcolor},
	tabsize=4,    
	language=ksh,
        basicstyle=\ttfamily\scriptsize,
        upquote=true,
        aboveskip={1.5\baselineskip},
        columns=fixed,
        showstringspaces=false,
        extendedchars=false,
        breaklines=true,
        prebreak = \raisebox{0ex}[0ex][0ex]{\ensuremath{\hookleftarrow}},
        frame=single,
        numbers=left,
        showtabs=false,
        showspaces=false,
        showstringspaces=false,
        identifierstyle=,
        keywordstyle=\color[rgb]{0,0,1},
        commentstyle=\color[rgb]{0.026,0.112,0.095},
        stringstyle=\color[rgb]{0.627,0.126,0.941},
        numberstyle=\color[rgb]{0.205, 0.142, 0.73},
}

\flushbottom 

\maketitle 

\tableofcontents


\thispagestyle{empty} 


\section{Introduction} 
Metis is a CCNx 1.0 forwarder written in C using the PARCLibrary package.  A forwarder is responsible
for receiving wire format packets from one place and forwarding them to another.  When a forwarder
runs on an end host, it typically forwards packets between applications, themselves, and between applications and
the network.  When a forwarder runs as an intermediate system, it typically forwarders between peers,
though it may have a small number of specialized applications, such as routing protocols, running
on the device.

This document describes the Metis architecture and principle data structures and algorithms.
Section~\ref{sec:architecture} provides a general architecture overview.  
Section~\ref{sec:usage} describes how to use Metis as a command line program
\daemon
and how to configure Metis with the command line program \control.
It also covers the syntax of a configuration file used by \daemon.
Section~\ref{sec:structure} describes the inner workings of Metis through
flow charts and key C structures.

\section{Architecture}\label{sec:architecture}
Metis is designed around the concept of a Connection as the atom of adjacency.  A Connection
can be a TCP or UDP connection (\{src\_ip, src\_port, dst\_ip, dst\_port\}), an Ethernet 
adjacency (\{smac, dmac, etherType\}), a UNIX domain socket connection, or an IP multicast
group (\{src\_ip, src\_port, group\_ip, group\_port\}).  The ConnectionTable tracks all these
adjacencies and provides a ConnectionID (CID).  The CID is used in other tables, such as
the forwarding table (FIB) to denote next (egress) hops and in the pending Interest table (PIT) to
denote previous (ingress) hop. 

The Metis forwarder is comprised of several major modules, the two principle ones being
the IO module and the message processor module.  The IO module consists of a set of
Listeners that implement the MetisListenerOps interface and a set of protocol Connections
(e.g. StreamConnection or UdpConnection) that implement the MetisIoOps interface.
Because each Connection is protocol specific, it can implement the correct \fname{send()}
function for the protocol and keep the protocol specific state it needs.

When Metis receives a packet, it converts the packet into a \fname{MetisMessage}, which is an extent (offset, length)
map of important TLV fields to their location inside a packet.  The \fname{MetisMessage} also carries information
about the ingress Connection.

The Message Processor receives all \fname{MetisMessage} and directs Interests and Content Objects to the
appropriate processing path.  The Message Processor encapsulates the Pending Interest Table (PIT), 
Forwarding Information Base (FIB), and Content Store (CS).
If a Content Object (in the form of a \fname{MetisMessage}) is returned to the ingress port, it is sent by calling the ingress connection's
\fname{send()} function.  If the Interest is to be forwarded, it is reference-count replicated to
each next hop's Connection \fname{send()} function.

\begin{figure}
\centering
\includegraphics[width=\linewidth,page=1,trim=0.25in 0.5in 0.25in 0.5in,clip=true]{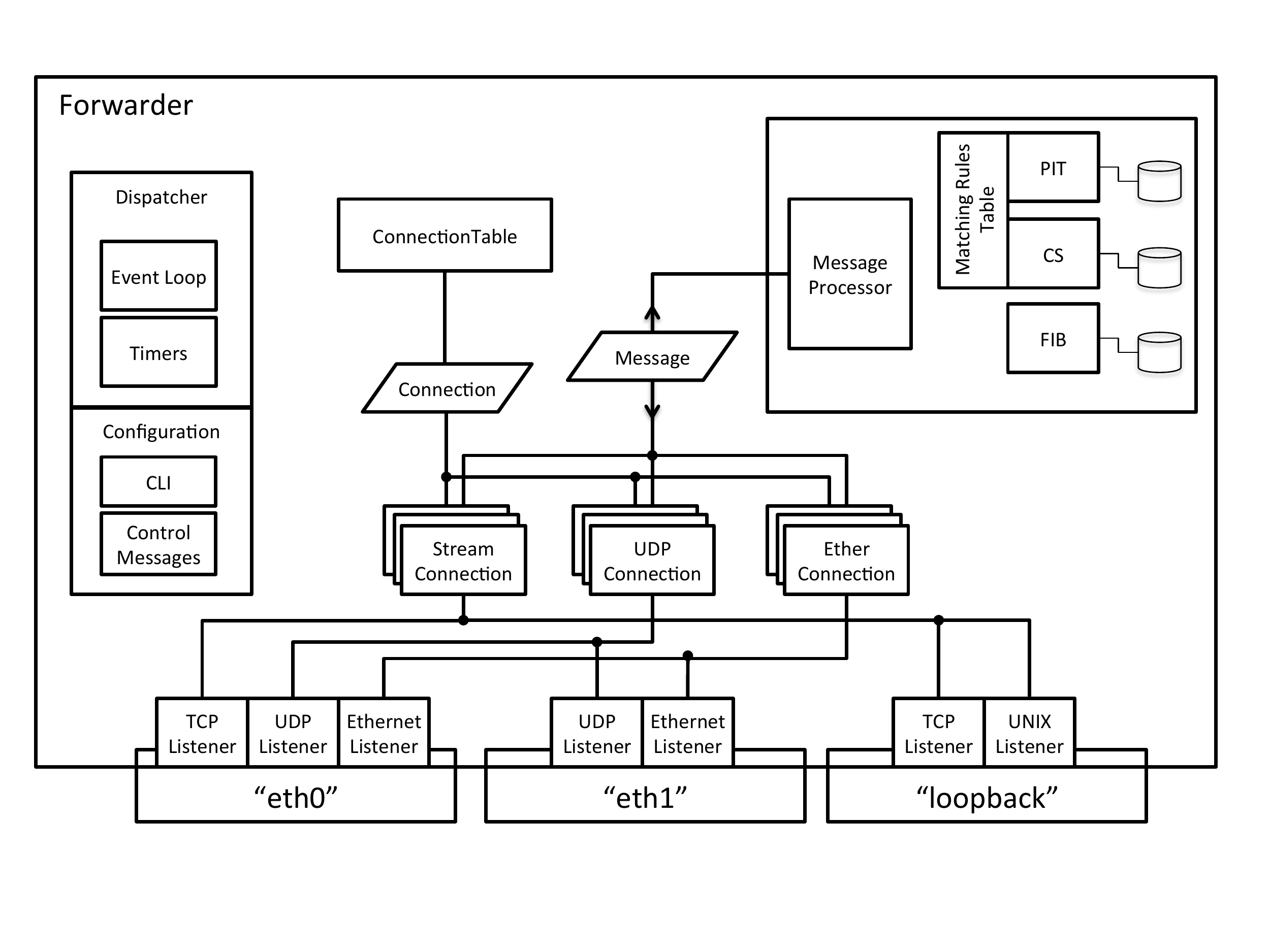}
\caption{Metis Architecture}
\label{fig:metis}
\end{figure}

A Content Store must comply with the \fname{MetisContentStoreInterface} (see Section~\ref{sec:contentstore}.
Metis provides a memory-backed transient Content Store implementation,
\fname{MetisLRUContentStore}, that uses an LRU replacement strategy.

Metis includes a \fname{MetisDispatcher} that is responsible for timers and socket polling.  It is a
single threaded, non-preemptive dispatcher.  Timers are serviced at the nearest time no earlier than
their expiration via a callback.  When a socket is readable or writeable, the dispatcher calls a
corresponding handler.  The current implementation uses \fname{PARCDispatcher}, which is based on Libevent.

A configuration module has a command parser used by both the configuration file and
to parse received configuration messages from the network.  Currently, the configuration messages
over the network use a JSON encoding. 

\subsection{Future Work}
While there is much yet left to do in Metis, these are some of the main future work items.

\subsubsection{Threading}
Metis will be threaded in a conventional ``reader - parser - lookup - writer'' model.  This will
generally correspond as reader and writer to the IO module, parser to \fname{MetisMessage} construction,
lookup to the message processor module, and writer to the IO module.

\subsubsection{Interface Generalization}
Not all interfaces are generalized to allow pluggable implementations.  In particular,
the FIB, CS, and Strategy sections still need work to bring up to a clean facade
pattern.

\subsubsection{Dispatcher and network I/O}
The current reliance on Libevent and the PARCDispatcher will be replaced with
a much leaner and properly generalized facade.  This will allow one to substitute
any suitable back-end for network IO.

\subsubsection{Ethernet}
There are plans on moving Ethernet to kernel bypass networking on Linux when 
supported by a backend like netmap/VALE or Intel DPDK.

The current Linux Ethernet, based on raw socket I/O, will be updated to use
shared kernel memory even without netmap or DPDK.  This is a small change
to introduce the shared memory kernel buffers.

The current Darwin Berkeley Packet Filter will be updated to use PF\_NDRV, which
should make it very similar to the current (non-shared memory) Linux raw socket module.

\subsubsection{Configuration Messages}
The current use of a proprietary fixed header PacketType and embedded JSON string
for a configuration message will be replaced with a CCNx 1.0 Control Message, which
is a signed Content Object.  We will be adding a certificate trust mechanism to Metis
along with ways to restrict which connections can receive control messages.

\section{Usage}\label{sec:usage}
This section describes how to run and configure Metis.  The content of this section
is the same as the man pages for metis\_daemon, metis\_control, and metis.cfg.

\subsection{Metis Daemon}
metis\_daemon ---  Metis is the CCNx 1.0 forwarder, which runs on each end system and as a software forwarder  on intermediate systems.   
\subsection*{Synopsis}
\label{metis-daemon-synopsis}\hyperlabel{synopsis}%

\texttt{met\penalty5000 i\penalty5000 s\penalty5000 \_\penalty5000 d\penalty5000 a\penalty5000 e\penalty5000 mon}  [-{}-{}port \texttt{\emph{\small{port}}}] [-{}-{}daemon ] [-{}-{}capacity \texttt{\emph{\small{con\penalty5000 t\penalty5000 e\penalty5000 n\penalty5000 t\penalty5000 S\penalty5000 t\penalty5000 o\penalty5000 r\penalty5000 e\penalty5000 S\penalty5000 ize}}}] [-{}-{}log \texttt{\emph{\small{fac\penalty5000 i\penalty5000 l\penalty5000 ity=\penalty0 level}}}...] [-{}-{}log-{}file \texttt{\emph{\small{log\penalty5000 f\penalty5000 ile}}}] [-{}-{}config \texttt{\emph{\small{con\penalty5000 f\penalty5000 i\penalty5000 g\penalty5000 f\penalty5000 ile}}}]

\subsection*{DESCRIPTION}
\label{metis-daemon-description}\hyperlabel{description}%

\textbf{metis\_daemon} is the CCNx 1.0 forwarder, which runs on each end system and as a software forwarder 
on intermediate systems.  metis\_daemon is the program to launch Metis, either as a console program 
or a background daemon (detatched from console).  Once running, use the program \emph{metis\_control} to 
configure Metis.

Metis is structured as a set of Listeners, each of which handles a specific method of listening for packets. 
For example, a TCP listener will accept connections on a specific TCP port on a specific local IP address. 
An Ethernet listener will accept frames of a specific EtherType on a specific Interface.

When Metis accepts a connection, it will create a Connection entry in the ConnectionTable to represent that peer. 
For Ethernet, a Connection is the tuple \{dmac, smac, ethertype\}.  For TCP and UDP, it is the tuple \{source IP, source port, 
destination IP, destination port\}.  The connid (connection ID) becomes the reverse route index in the Pending Interest Table.

\subsection*{OPTIONS}
\label{metis-daemon-options}\hyperlabel{options}%

\noindent
\begin{description}
\item[{-{}-{}config \emph{configfile}}] Reads configuration parameters from 
\emph{configfile}. 
The  
\texttt{-{}\penalty0 -{}\penalty0 port} option has no effect in this mode 
and Metis will not listen to any ports.  This means that  
\emph{metis\_control} will not be able to connect to Metis to configure it 
further unless one includes at least a listener for TCP localhost or a unix domain socket.
\item[{-{}-{}capacity \emph{contentStoreSize}}] Sets the capacity of the Content Store to  
\emph{contentStoreSize} content objects. 
Metis uses a least-{}recently-{}used eviction policy.  A size of 0 will disable the 
Content Store.

The Content Store sits on the fast path of the forwarder, so there is a cost 
associated with adding and removing items to the Content Store tables.
\item[{-{}-{}daemon}] Runs Metis in daemon mode, detaching from the console.  It must 
be run with the \texttt{-{}\penalty0 -{}\penalty0 log-{}\penalty0 file} option.
\item[{-{}-{}log \emph{facility}=\emph{level}}] Sets the log level of the given  
\emph{facility} 			to the given 
\emph{level}. 
The \texttt{-{}\penalty0 -{}\penalty0 log} option may be repeated 
several times setting the log level of different facilities.  If the same 
facility is listed twice, only the last occurance takes effect. 
The default log level is Error for all facilities.

Facilities:

\noindent
\begin{description}
\item[{   \textbullet{}}] all: All facilities.
\item[{   \textbullet{}}] config: Configuration activies.
\item[{   \textbullet{}}] core: Core forwarder, such as startup and shutdown.
\item[{   \textbullet{}}] io: Listeners, connections, and all I/O related activities.
\item[{   \textbullet{}}] message: CCNx messages, such as parsing.
\item[{   \textbullet{}}] processor: Forwarding processor, such as CS, FIB, and PIT activities.
\end{description}

The log levels are: debug, info, notice, warning, error, critical, alert, off.
\item[{-{}-{}log-{}file \nolinkurl{logfile}}] Specifies the 
\emph{logfile} 			to write all log messages.  This parameter is required with  
\texttt{-{}\penalty0 -{}\penalty0 dae\penalty5000 mon} mode.
\item[{-{}-{}port \emph{port}}] The UDP and TCP port to listen on.  If no  
\emph{configfile}             is specified, Metis will listen on this port on all interfaces 
including localhost.

If this parameter is not given, Metis uses the default port 9695.
\end{description}

\subsection*{USAGE}
\label{metis-daemon-usage}\hyperlabel{usage}%

\textbf{metis\_daemon} -{}-{}config metis.cfg -{}-{}log all=info -{}-{}log config=debug -{}-{}log-{}file metis.log

\subsection*{SEE ALSO}
\label{metis-daemon-see-also}\hyperlabel{see-also}%

See \emph{metis\_control}\emph{(1)} for a 
description of how to configure \textbf{metis\_daemon}.

For a list of all configuration lines that may be used with 
\emph{metis\_control} and by \texttt{-{}\penalty0 -{}\penalty0 con\penalty5000 fig} configuration file, 
see \emph{metis.cfg}\emph{(5)}.

\subsection*{CAVEATS}
\label{metis-daemon-caveats}\hyperlabel{caveats}%

\noindent
\begin{description}
\item[{   \textbullet{}}] A given interface may only have one Ethernet listener on one EtherType.
\item[{   \textbullet{}}] If there are multiple longest matching prefix entries that match an Interest, it will be 
forwarded to all those routes (i.e. multicast).
\item[{   \textbullet{}}] Ethernet fragmentation will only use the interface MTU and there is no MTU discovery.  If Metis is 
used in a bridged environment, this may lead to errors if the MTU changes on different segments, such 
as a 10G link at 9000 bytes and a 100 Mbps link at 1500 bytes.
\end{description}

\subsection{Metis Control}
metis\_control ---  Metis is the CCNx 1.0 forwarder, which runs on each end system and as a software forwarder  on intermediate systems.  metis\_control is the program to configure the forwarder,   metis\_daemon.   
\subsection*{Synopsis}
\label{metis-control-synopsis}\hyperlabel{synopsis}%

\texttt{met\penalty5000 i\penalty5000 s\penalty5000 \_\penalty5000 c\penalty5000 o\penalty5000 n\penalty5000 t\penalty5000 rol}  [-{}-{}keystore \texttt{\emph{\small{key\penalty5000 s\penalty5000 t\penalty5000 ore}}}] [-{}-{}password \texttt{\emph{\small{pas\penalty5000 s\penalty5000 w\penalty5000 ord}}}] [\texttt{\emph{\small{com\penalty5000 m\penalty5000 a\penalty5000 n\penalty5000 d\penalty5000 l\penalty5000 ine}}}]

\subsection*{DESCRIPTION}
\label{metis-control-description}\hyperlabel{description}%

\textbf{metis\_control} 	is the program used to configure a running forwarder \emph{metis\_daemon}.  It will connect to 
the forwarder over a local listener (e.g. TCP to localhost or a unix domain socket).  If a  
\emph{commandline} option is specified, \textbf{metis\_control} 	will send that one command to Metis and then exit.  If no \emph{commandline} 	is specified, \emph{metis\_command} will enter interacitve mode where the user can issue 
multiple commands.

\textbf{metis\_control} requires a signing keystore for communicating over the network.  The 
\emph{keystore} file is a standard PKCS12 keystore, and may be 
created using 
\emph{parc\_public\-key}\emph{(1)}. 
If no \emph{keystore} is specified, \textbf{metis\_control} 	will look in the standard path \textasciitilde{}/.ccnx/.ccnx\_keystore.p12. 
The keystore password is specified in \emph{password}.  If not specified, 
no password is used.  If the keystore does not open, the user will be prompted for a password.

See \emph{metis.cfg}\emph{(5)} for 
a specification of the available \emph{commandline}.

The environment variable METIS\_PORT may be used to specify what TCP port to use to connect to the local Metis. 
The environment variable METIS\_LOCALPATH may be used to specify the UNIX domain socket to connect to the local Metis 
and takes priority over METIS\_PORT.

\subsection*{OPTIONS}
\label{metis-control-options}\hyperlabel{options}%

\noindent
\begin{description}
\item[\breakitem{-{}-{}keystore \emph{keystore}}] \textbf{metis\_control} requires a signing keystore for communicating over the network.  The 
\emph{keystore} file is a standard PKCS12 keystore, and may be 
created using 
\emph{parc\_publickey}\emph{(1)}. 
If no \emph{keystore} is specified, \textbf{metis\_control} 				will look in the standard path \textasciitilde{}/.ccnx/.ccnx\_keystore.p12.
\item[\breakitem{-{}-{}password \emph{password}}] The keystore password is specified in \emph{password}.  If not specified, 
no password is used.  If the keystore does not open, the user will be prompted for a password.
\item[{commandline}] The remainder of the arguments are the commandline to send to Metis.  See USAGE.
\end{description}

\subsection*{USAGE}
\label{metis-control-usage}\hyperlabel{usage}%

\textbf{metis\_control} -{}-{}keystore keystore.p12

\textbf{metis\_control} -{}-{}keystore keystore.p12 list interfaces

\subsection*{SEE ALSO}
\label{metis-control-see-also}\hyperlabel{see_also}%

See \emph{parc\_publickey}\emph{(1)} for a utility 
to create a PKCS keystore.

For a list of all configuration lines that may be used with 
\textbf{metis\_control} and by \texttt{-{}\penalty0 -{}\penalty0 con\penalty5000 fig} configuration file, 
see \emph{metis.cfg}\emph{(5)}.

The default keystore is \textasciitilde{}/.ccnx/.ccnx\_keystore.p12.

\subsection{Metis Configuration File}

metis.cfg is an example of a configuation file usable with  metis\_daemon(1),  though there is nothing special about the actual filename.  Each line of the configuration file is also usable with   metis\_control(1).  This  document specifies all available command lines used to configure and query Metis.    All commands have a 'help', so typing 'help command' will display on-{}line help.     In a configuration file, lines beginning with '\#' are comments.   
\subsection*{ADD COMMANDS}
\label{metis-cfg-add-commands}

\noindent
\begin{description}
\item[{add connection ether \emph{symbolic} \emph{dmac} \emph{interface}}] Adds an Ethernet connection on \emph{interface}  to the given destination MAC address. 
The \emph{symbolic} name is a symbolic name for the connection, which may be used in 
later commands, such as \emph{add route}. 
There must be an Ethernet Listener on the specified interface (see \emph{add listener}), and the connection  
will use the same EtherType as the Listener. The \emph{dmac} destination MAC address 
is in hexidecimal with optional "-{}" or ":" separators.

A connection is a target for a later route assignment or for use as an ingress identifier in the PIT.  When using a broadcast 
or group address for a connection, an Interest routed over that connection will be broadcast.  Many receivers may respond. 
When Metis receives a broadcast Interest it uses the unicast source MAC for the reverse route -{}-{} it will automatically create 
a new connection for the source node and put that in the PIT entry, so a Content Object answering the broadcast Interest will 
only be unicast to the previous hop.

add connection ether conn7 e8-{}06-{}88-{}cd-{}28-{}de em3

add connection ether bcast0 FFFFFFFFFFFF eth0
\item[{\breakitem{add connection (tcp | udp) \emph{symbolic} \emph{remote\_ip} \emph{remote\_port} \emph{local\_ip} \emph{local\_port}}}] Opens a connection to the specific \emph{remote\_ip} (which may be a hostname, though you do not have control over IPv4 or IPv6 in this case) on \emph{remote\_port}.  The local endpoint is given by \emph{local\_ip} \emph{local\_port}.  While the \emph{local\_ip} \emph{local\_port} are technically optional parameters, the system's choice of local address may not be what one expects or may be a different protocols (4 or 6).  The default port is 9695.

A TCP connection will go through a TCP connection establishment and will not register as UP until the remote side accepts.  If one side goes down, the TCP connection will not auto-{}restart if it becomes availble again.

A UDP connection will start in the UP state and will not go DOWN unless there is a serious network error.

\noindent
\begin{description}
\item[\breakitem{Opens a connection to 1.1.1.1 on port 1200 from the local address 2.2.2.2 port 1300}] add connection tcp conn0 1.1.1.1 1200 2.2.2.2 1300
\item[\breakitem{opens connection to IPv6 address on port 1300}] add connection udp barney2 fe80::aa20:66ff:fe00:314a 1300
\end{description}
\item[\breakitem{add listener (tcp|udp) \emph{symbolic} \emph{ip\_address} \emph{port}, add listener ether \emph{symbolic} \emph{interfaceName} \emph{ethertype}, add listener local \emph{symbolic} \emph{path}}] Adds a protocol listener to accept packets of a given protocol (TCP or UDP or Ethernet). 
The \emph{symbolic} name represents the listener and will be used in future commands 
such as access list restrictions.  If using a configuration file on \emph{metis\_daemon}, you must include 
a listener on localhost for local applications to use.

The \emph{ip\_address} is the IPv4 or IPv6 local address to bind to. 
The \emph{port} is the TCP or UDP port to bind to.

The \emph{interfaceName} is the interface to open a raw socket on (e.g. "eth0"). 
The \emph{ethertype} is the EtherType to use, represented as a 0x hex number (e.g. 0x0801) 
or an integer (e.g. 2049).

The \emph{path} parameter specifies the file path to a unix domain socket.  Metis  
will create this file and remove it when it exits.

\noindent
\begin{description}
\item[\breakitem{Listens to 192.168.1.7 on tcp port 9695 with a symbolic name 'homenet'}] add listener tcp homenet 192.168.1.7 9695
\item[\breakitem{Listens to IPv6 localhost on udp port 9695}] add listener udp localhost6 ::1 9695
\item[\breakitem{Listens to interface 'en0' on ethertype 0x0801}] add listener ether nic0 en0 0x0801
\end{description}
\item[\breakitem{add route \emph{symbolic} \emph{prefix} \emph{prefix}}] Adds a static route to a given \emph{prefix} to the FIB for longest match.

Currently, the \emph{symbolic} and \emph{cost} are not used.
\end{description}

\subsection*{LIST COMMANDS}
\label{metis-cfg-list-commands}%

\noindent
\begin{description}
\item[\breakitem{list connections}] Enumerates the current connections to Metis.  These include all TCP, UDP, Unix Domain, and Ethernet peers. 
Each connection has an connection ID (connid) and a state (UP or DOWN) followed by the local (to metis) and remote 
addresses.
\item[\breakitem{list interfaces}] Enumerates the system interfaces available to Metis.  Each interface has an Interface ID, a 'name' (e.g. 'eth0'), 
an MTU as reported by the system, and one or more addresses.
\item[{list routes}] Enumerates the routes installed in the FIB.  
The \emph{iface} is the out-{}bound connection.   
The \emph{protocol} is the the routing protocol that injected the route. 
'STATIC' means it was manually entered via \emph{metis\_control}. 
\emph{route} is the route type.  'LONGEST' means longest matching prefix 
and 'EXACT' means exact match.  Only 'LONGEST' is supported. 
\emph{cost} is the cost of the route.  It is not used. 
\emph{next} is the nexthop on a multiple access interface.  it is not used 
because the current implementation uses one connection (iface)  per neighbor. 
\emph{prefix} is the CCNx name prefix for the route.
\item[{Examples}] ~
\begin{lstlisting}[firstnumber=1,language=ksh]
> list connections 
23   UP inet4://127.0.0.1:9695 inet4://127.0.0.1:64260 TCP 
 
> list interfaces 
int       name lm      MTU  
24        lo0 lm    16384 inet6://[::1\%0]:0 
inet4://127.0.0.1:0 
inet6://[fe80::1\%1]:0 
25        en0  m     1500 link://3c-15-c2-e7-c5-ca 
inet6://[fe80::3e15:c2ff:fee7:c5ca\%4]:0 
inet4://13.1.110.60:0 
inet6://[2620::2e80:a015:3e15:c2ff:fee7:c5ca\%0]:0 
inet6://[2620::2e80:a015:a4b2:7e10:61d1:8d97\%0]:0 
26        en1  m     1500 link://72-00-04-43-4e-50 
inet4://192.168.1.1:0 
27        en2  m     1500 link://72-00-04-43-4e-51 
28    bridge0  m     1500 link://3e-15-c2-7e-96-00 
29       p2p0  m     2304 link://0e-15-c2-e7-c5-ca 
 
> list routes 
iface  protocol   route     cost                 next prefix 
23    STATIC LONGEST        1 ---.---.---.---/.... lci:/foo/bar 
Done 
 
\end{lstlisting}
\end{description}

\subsection*{REMOVE COMMANDS}
\label{metis-cfg-remove-commands}%

\noindent
\begin{description}
\item[{remove connection}] Not implemented.
\item[{remove route}] Not implemented.
\end{description}

\subsection*{MISC COMMANDS}
\label{metis-cfg-misc-commands}%

\noindent
\begin{description}
\item[{quit}] In interactive mode of \emph{metis\_control}, it cause the program to exit.
\item[{set debug}] Turns on the debugging flag in \emph{metis\_control} to display information about its connection to Metis.
\item[{unset debug}] Turns off the debugging flag in \emph{metis\_control} to display information about its connection to Metis.
\end{description}

\subsection*{USAGE}
\label{metis-cfg-usage}%

\emph{Example Linux metis.cfg configuration file}

\begin{lstlisting}[firstnumber=1,language=ksh]
#local listeners for applications 
add listener tcp local0 127.0.0.1 9695 
add listener udp local1 127.0.0.1 9695 
add listener local unix0 /tmp/metis.sock 
 
# add ethernet listener and connection 
add listener ether nic0 eth0 0x0801 
add connection ether conn0 ff:ff:ff:ff:ff:ff eth0 
add route conn0 lci:/ 1 
 
# add UDP tunnel to remote system 
add connection udp conn1 ccnx.example.com 9695 
add route conn1 lci:/eample.com 1 
 
\end{lstlisting}

\emph{Example one-{}shot metis\_control commands}

\begin{lstlisting}[firstnumber=1,backgroundcolor={},basicstyle=\ttfamily,escapeinside={<t>}{</t>},moredelim={**[is][\bfseries]{<b>}{</b>}},moredelim={**[is][\itshape]{<i>}{</i>}},]
<i>metis_control</i> list routes 
<i>metis_control</i> add listener local unix0 /tmp/metis.sock 
 
\end{lstlisting}

\section{Internal Structure}\label{sec:structure}
... put stuff here ...

\subsection{Connection State Machine}
A Connection (see below, Section~\ref{sec:connection}) follow this state machine:

{\small
\begin{verbatim}
   initial   -> CREATE
   CREATE    -> (UP | DOWN)
   UP        -> (DOWN | DESTROYED)
   DOWN      -> (UP | CLOSED | DESTROYED)
   CLOSED    -> DESTROYED
   DESTROYED -> terminal
\end{verbatim}}

These states should be signaled via the \fname{MetisMessenger} (see Section~\ref{sec:messenger}
to any component that wishes to subscribe to connection event messages.
It is the responsibility of the Listener (Section~\ref{sec:listeners}) and IO Connection
(Section~\ref{sec:io-connection}) to generate these signals.

\subsection{Messanger}\label{sec:messenger}
The \fname{MetisMessenger} interface inside Metis is to send internal signals of events.  A module can
subscribe to receive messages via \fname{metisMessenger\_Register()}.  When any component signals
a message via \fname{metisMessenger\_Send()}, all \fname{MetisMessengerRecipient} callbacks will
receive the message \emph{in a later dispatcher scheduling time}.

The essential element of the Messenger is the  \emph{in a later dispatcher scheduling time} condition.
This avoids pre-emption and circular callback firing.

Figure~\ref{api:messenger} shows the interesting API functions of the messenger.  Currently,
a \fname{MetisMissive} can only signal the state machine for a Connection ID.

\begin{figure}
\begin{lstlisting}
void metisMessenger_Send(MetisMessenger *messenger, MetisMissive *missive);
void metisMessenger_Register(MetisMessenger *messenger, const MetisMessengerRecipient *recipient);
void metisMessenger_Unregister(MetisMessenger *messenger, const MetisMessengerRecipient *recipient);
\end{lstlisting}
\caption{MetisMessenger API}
\label{api:messenger}
\end{figure}

\subsection{Configuration}
FINISH

\subsection{Listeners}\label{sec:listeners}
When Metis starts up, it will either create a set of default listeners (TCP, UDP) on a given port
or only create those listeners specified in the configuration file.  All listeners implement the \fname{MetisListenerOps}
interface (see Figure~\ref{struct:MetisListenerOps}).

The job of a listener is to receive a packet from the network, associate it with a \fname{MetisConnection}, 
create a \fname{MetisMessage}, and send it to the \fname{MetisMessageProcessor}.
For stream listeners, the \fname{accept()} happens in the listener, and from then on the per-client
socket IO happens in \fname{MetisStreamConnection}.  For datagram listeners (UDP and Ethernet), the
Listener has to do all the initial IO to at least match against a Connection.  In the current code,
the Listener does all the IO -- matching to a Connection and creating the \fname{MetisMessage}.

\begin{figure}
\begin{lstlisting}
struct metis_listener_ops {
    void *context;
    void (*destroy)(MetisListenerOps **listenerOpsPtr);
    unsigned (*getInterfaceIndex)(const MetisListenerOps *ops);
    const CPIAddress * (*getListenAddress)(const MetisListenerOps *ops);
    MetisEncapType (*getEncapType)(const MetisListenerOps *ops);
    int (*getSocket)(const MetisListenerOps *ops);
};
\end{lstlisting}
\caption{MetiisListenerOps}
\label{struct:MetisListenerOps}
\end{figure}

The \fname{destroy()} function is called during cleanup to release the listener.  The
\fname{getInterfaceIndex()} function returns which host interface the listener is bound to.
The \fname{getListenAddress} is the host address the listener is bound to.
The \fname{getEncapType()} function is used to display listener information and
indicates the encapsulate (TCP, Ethernet, etc.) used by the listener.
The \fname{getSocket()} function is used by some protocol connections when they
need to send a packet from the listeners socket address, such as UDP.

In the case of Ethernet, the Listener is split between a platform-specific module for the low-level IO
module called \fname{MetisGenericEther} (see Figure~\ref{struct:MetisGenericEther}) and
the high-level \fname{MetisListener}.  \fname{MetisGenericEther} is the header that each
platform-specific Ethernet module implements.  It is not a structure-style facade, but a straight
header as we expect only one platform-specific object file per platform.

\begin{figure}
\begin{lstlisting}[language=C]
MetisGenericEther *metisGenericEther_Create(MetisForwarder *metis, const char *deviceName, uint16_t etherType);
MetisGenericEther *metisGenericEther_Acquire(const MetisGenericEther *ether);
void metisGenericEther_Release(MetisGenericEther **etherPtr);
int metisGenericEther_GetDescriptor(const MetisGenericEther *ether);
bool metisGenericEther_ReadNextFrame(MetisGenericEther *ether, PARCEventBuffer *buffer);
bool metisGenericEther_SendFrame(MetisGenericEther *ether, PARCEventBuffer *buffer);
PARCBuffer *metisGenericEther_GetMacAddress(const MetisGenericEther *ether);
uint16_t metisGenericEther_GetEtherType(const MetisGenericEther *ether);
unsigned metisGenericEther_GetMTU(const MetisGenericEther *ether);
\end{lstlisting}
\caption{MetisGenericEther platform-specific interface}
\label{struct:MetisGenericEther}
\end{figure}

\subsubsection{Stream Listeners (TCP, Unix)}
\begin{figure}[th]
\centering
\includegraphics[width=\linewidth,trim=0in 0in 0in 0in,clip=true]{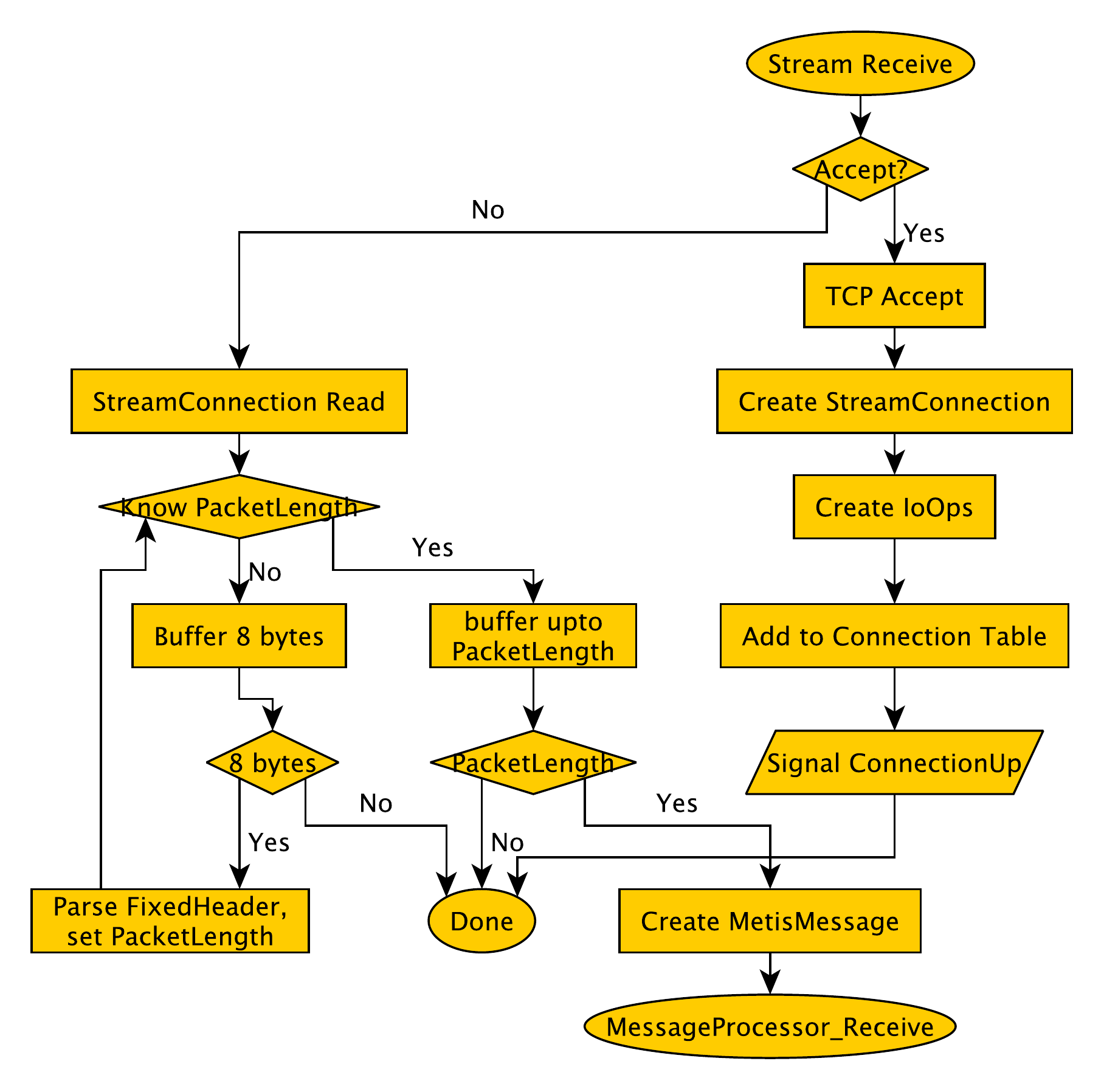}
\caption{Stream Receive}
\label{fig:StreamReceive}
\end{figure}

Figure~\ref{fig:StreamReceive} shows the process of \fname{TcpListener}, \fname{UnixListener}, and
\fname{StreamConnection} when receiving
a packet.  Because it is a stream connection, we must do our own framing based on the Fixed Header.
\fname{StreamConnection} currently does not have any framing error recovery.
\fname{TcpListener} and \fname{UnixListener} are invoked to accept a new connection, and go through
the (for example) TCP Accept process.  This creates a \fname{StreamConnection} and associates it with
the client socket, creates the \fname{MetisIoOps} associated with TCP, and adds it to the Connection Table.
Once the connection is ready to go, it also sends a Metis Messenger signal that the connection is in the UP state.

Inside \fname{StreamConnection}, we need to maintain state about framing because bytes may arrive
with arbitrary delineation not corresponding to CCNx 1.0 packets. 
If we do not know the PacketLength, then we have not read a Fixed Header yet.  We buffer until we
have read 8 bytes and can parse the Fixed Header.
Once we know the PacketLength from
a Fixed Header, then we read the socket up to PacketLength bytes or the end of the available bytes (non-blocking).
Once we have read PacketLength bytes, we can create a \fname{MetisMessage} from the buffer and
pass it to the Message Processor.

\subsubsection{UDP Listener}
\begin{figure}[th]
\centering
\includegraphics[width=\linewidth,trim=0in 0in 0in 0in,clip=true]{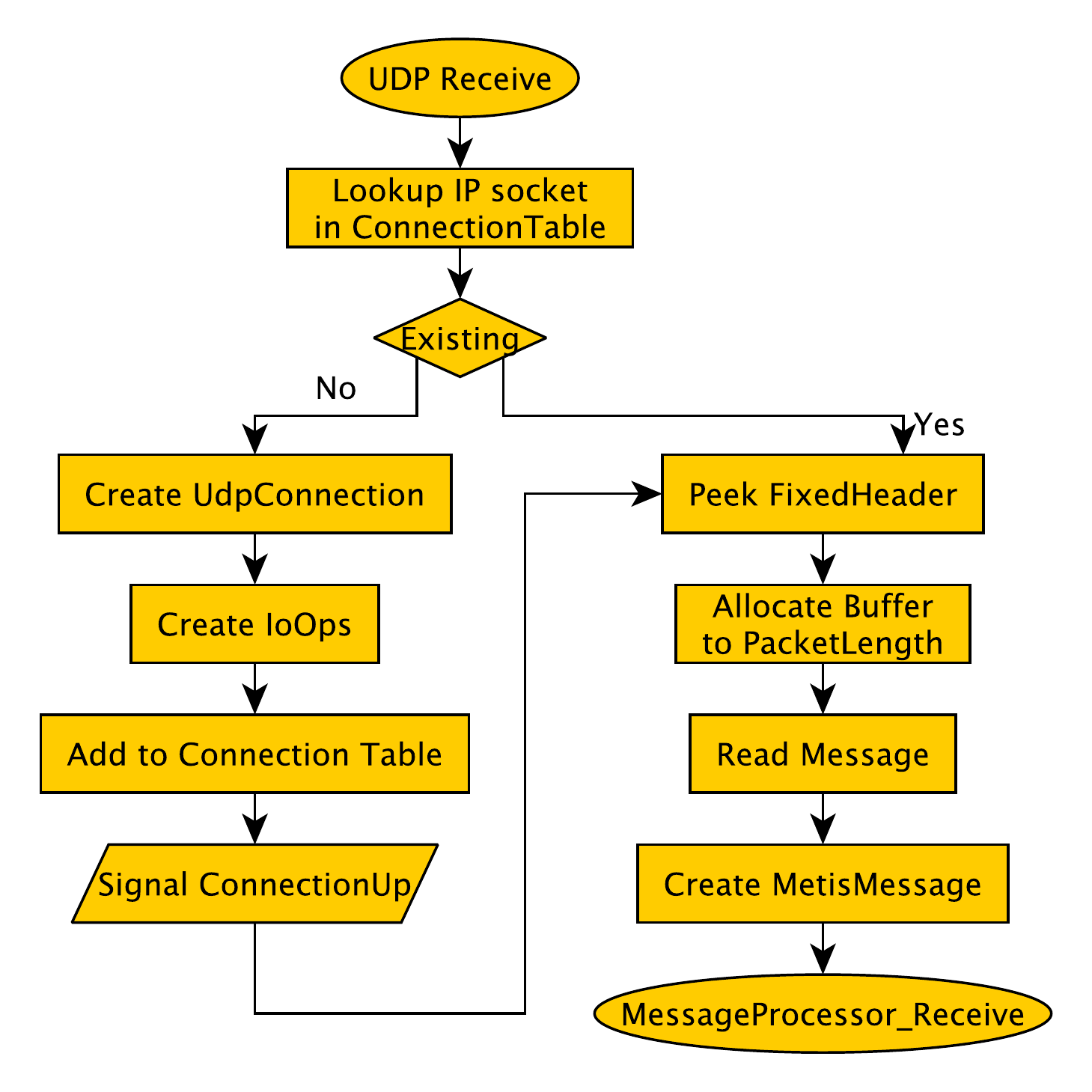}
\caption{UDP Receive}
\label{fig:UdpReceive}
\end{figure}

Figure~\ref{fig:UdpReceive} shows the process of \fname{UdpListener} receiving a packet.
Because UDP is datagram based, we do not need to manage framing as in the \fname{StreamConnection}.
However, as there is no dedicated client socket, the \fname{UdpListener} must construct a
key for the Connection Table from the source and destination socket addresses to lookup
(or create) a corresponding Connection.  Creating a connection is the same as previously
described for TCP, except the \fname{MetisIoOps} concrete class is \fname{MetisUdpConnection}.
The UDP receive process currently does not have a buffer pool, so it peeks at the FixedHeader
bytes to determine how bit a buffer to allocate and then reads the packet in to that buffer.
This process is inefficient because it requires two system calls per read.
Once the packet is read, we proceed as above creating a \fname{MetisMessage} and
passing it to the Message Processor.

\subsubsection{Ethernet Listener}
\begin{figure}[th]
\centering
\includegraphics[width=\linewidth,trim=0in 0in 0in 0in,clip=true]{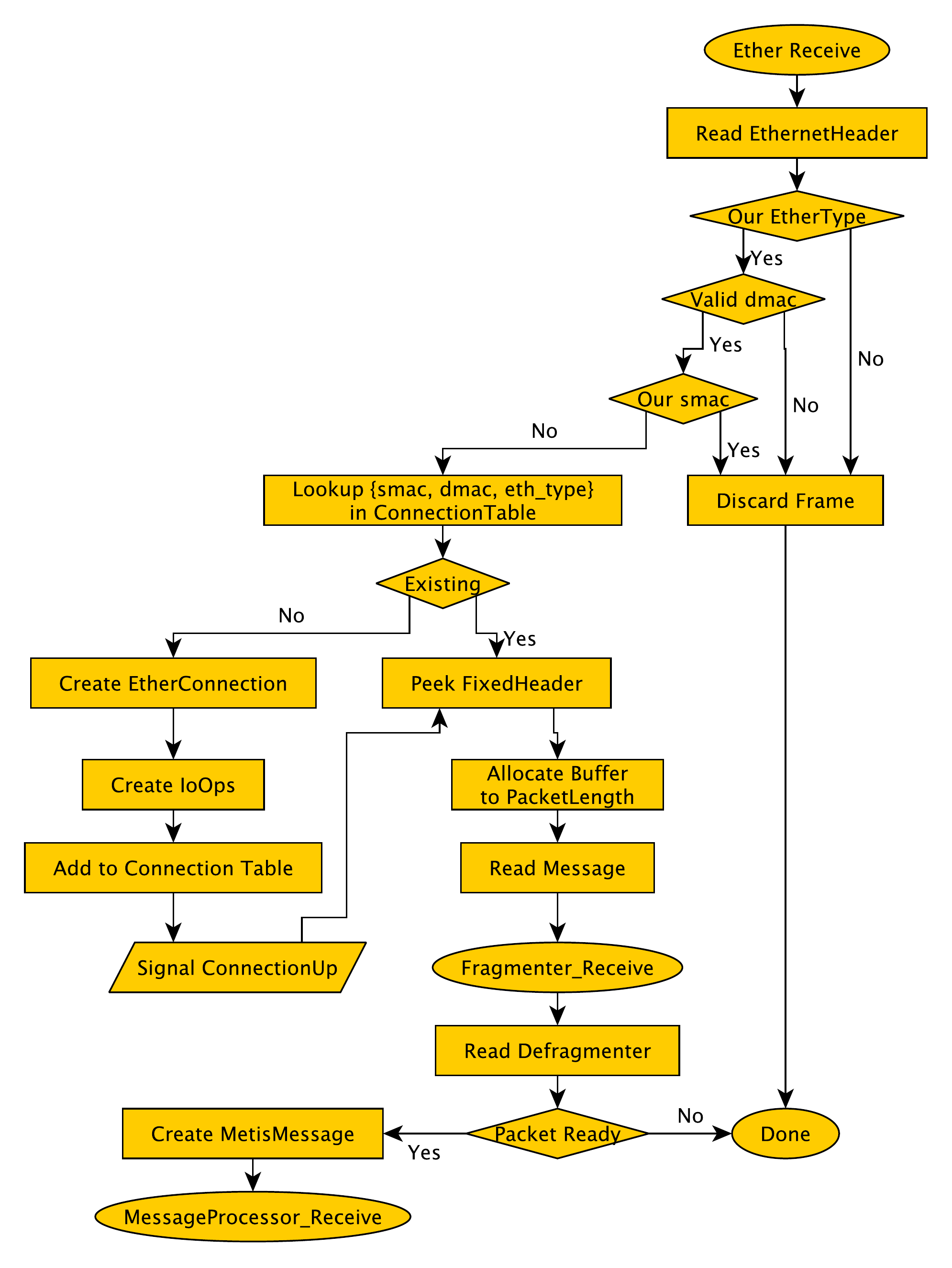}
\caption{Ethernet Receive}
\label{fig:EtherReceive}
\end{figure}

Figure~\ref{fig:EtherReceive} shows the Ethernet receive process down to the \fname{GenericEther}
abstraction level, which does not include the low-level platform specific parts.  These will differ
between linux and Mac and other paltforms.  The platform Ethernet implementation may need
to trim the CRC from the packet, as some platforms strip it and some do not.

The Ethernet process is similar to the UDP process
in that there is no client socket, so the Ethernet listener needs to resolve the Connection by
doing its own query to the Connection Table.  

The first steps are to ensure the received Ethernet frame matches our EtherType, an acceptable
destination MAC address (dmac) and is not our source MAC address (smac).  Acceptable
dmac addresses include the interface hardware address, the broadcast address, and the CCNx
Ethernet group address.  If the packet passes these tests, we lookup the address tuple
\{smac, dmac, etherType\} in the Connection Table and create a new \fname{MetisEtherConnection}
if needed.  Creating a new \fname{MetisIoOps} proceeds as above.  

Once we are past the
Ethernet header, we can read the Fixed Header, allocate a buffer for the \fname{MetisMessage}  and
read the packet in to that buffer.  The exact memory mechanics that happen here can vary depending
on the platform Ethernet implementation.  Once we have a \fname{MetisMessage} , it is passed to
the Message Processor.

\subsection{IO Connections}\label{sec:io-connection}
An IO Connection is code that implements the \ioops interface, shown in Figure~\ref{struct:MetisIoOps}.
Each \ioops represents a connection, so it has a \fname{getAddressPair()} function.  The addresses are of
type \fname{CPIAddress} which holds IP and UNIX and Ethernet addresses. 

The \fname{send()} function is used by the Message Process to send a Content Object back along an Interest reverse path
and to forward an Interest to next hops in the FIB. 

The \fname{isUp()} function indicates if the connection is able to send packets.  Sometimes a connection is valid, but
is not up.  For example a TCP connection will be valid but not Up during the time it is connecting to a remote peer.

The \fname{isLocal()} function indicates if the remote address is local to the host.   Ethernet is never local.  IP
addresses to the IPv4 and IPv6 loopback address are always local.  UNIX domain sockets are always local.

The \fname{getConnectionId()} function returns an integer representing the connection.  It may be used as a foreign key
in other tables.

The \fname{destroy()} function will release the connections memory.

The \fname{class()} function returns a unique \fname{void *} for the connection that represents the underlying
protocol.  It is used by function like 
\fname{metisEtherConnecion\_IsInstanceOf()}
to determine if a connection is of a particular type.

\begin{figure}
\begin{lstlisting}[language=C]
struct metis_io_ops {
    void *closure;
    bool (*send)(MetisIoOperations *ops, const CPIAddress *nexthop, MetisMessage *message);
    const CPIAddress *       (*getRemoteAddress)(const MetisIoOperations *ops);
    const MetisAddressPair * (*getAddressPair)(const MetisIoOperations *ops);
    bool (*isUp)(const MetisIoOperations *ops);
    bool (*isLocal)(const MetisIoOperations *ops);
    unsigned (*getConnectionId)(const MetisIoOperations *ops);
    void (*destroy)(MetisIoOperations **opsPtr);
    const void * (*class)(const MetisIoOperations *ops);
    CPIConnectionType (*getConnectionType)(const MetisIoOperations *ops);
};
\end{lstlisting}
\caption{MetisIoOps}
\label{struct:MetisIoOps}
\end{figure}

\subsection{MetisConnection}\label{sec:connection}
A \fname{MetisConnection} is a PARC-style object that encapsulates a \ioops for
storage in the Connection Table.  It supports the common functions like \fname{acquire()}
and \fname{release()}.  Other tables store the Connection ID instead of a reference to
a Connection.  This allows the connection to be taken down or removed without
needing to flush all other objects that reference the Connection, such as \fname{MetisMessage}
and FIB entries.

If a connection is removed while there are still references to its connection ID in the
system, they will be lazily purged when they try to reference the connection ID in
the connection table.

\subsection{Connection Table}
The ConnectionTable stores the state of every connection known to Metis.  These include
configured connections and tunnels (connections to remote systems) and ephemerally learned
connections such as receiving a UDP or Ethernet packet.

Ephemeral connections will timeout.  TCP connections automatically timeout when the TCP
session ends, as that causes a socket error that causes the connection to go to DOWN then CLOSED
state and the Connection Manager will remove it.  UDP and Ethernet connections need to manage
their own timeout and eventually go to DOWN and CLOSED state to be removed from the Connection Table.

\emph{Currently, UDP and Ethernet connections are not timing out.}

\subsection{Message Processor}
\begin{figure}
\centering
\includegraphics[width=\linewidth,trim=0in 0in 0in 0in,clip=true]{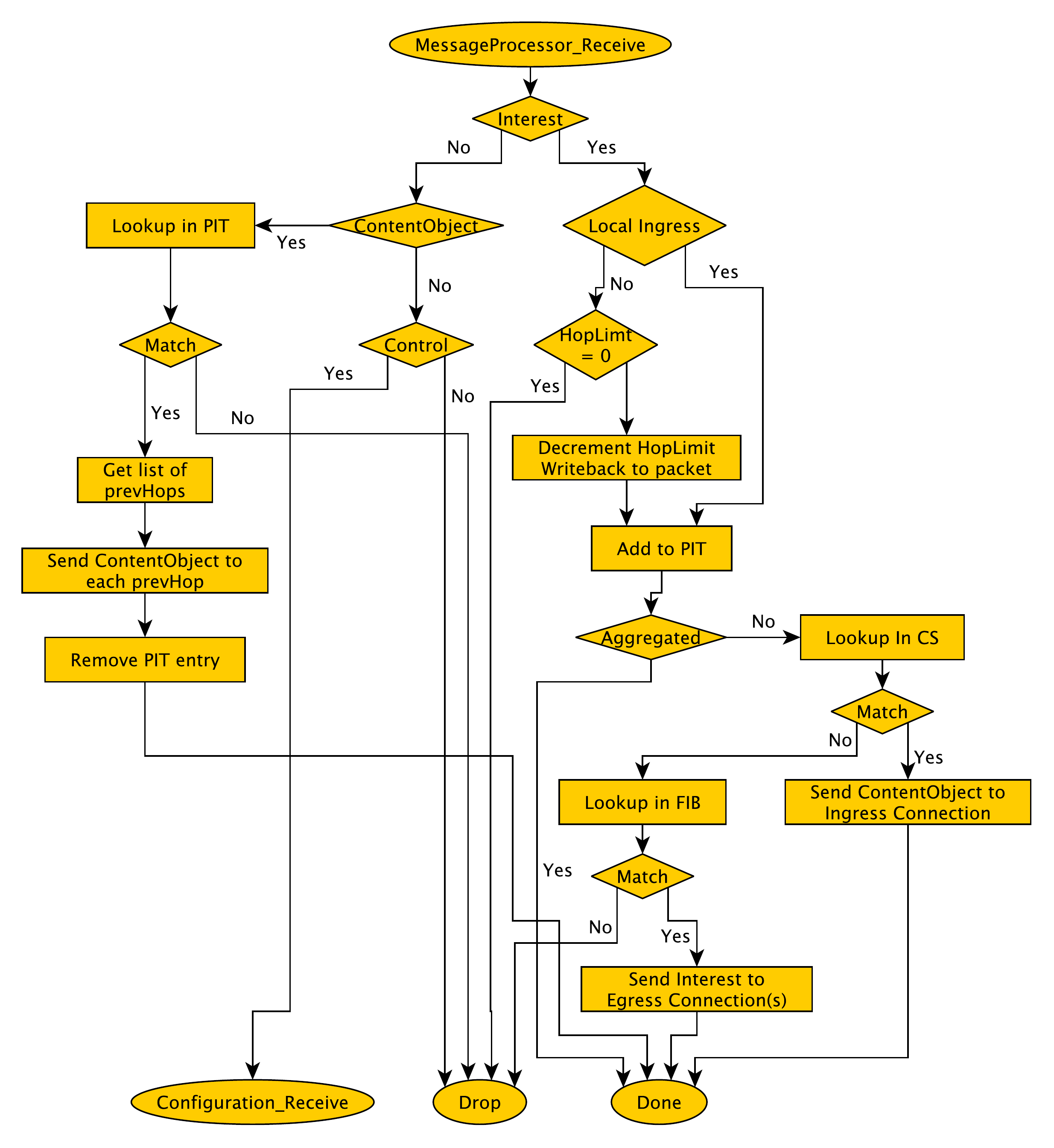}
\caption{Message Processor Receive}
\label{fig:MessageProcessorReceive}
\end{figure}

The Message Processor has an Interest and a Content Object processing path.  These paths
execute the normal CCNx 1.0 algorithm for each message type.  Figure~\ref{fig:MessageProcessorReceive}
shows the two processing paths.

An Interest message carries a HopLimit, which must be decremented if received from a remote system.
If the Interest is from a local application, the HopLimit is not decremented on receive.  If an interest
is aggregated in the PIT, then the message processor is done.  If the message is not aggregated --
it's a new Interest or the PIT determines it should be forwarded anyway -- then the Message Processor
tries to satisfy from the Content Store (if configured), and the tries to forward via the FIB.

If an Interest is satisfied from the Content Store, the corresponding Content Object is sent to the
ingress Connection's \fname{send()} function.  If the interest is forwarded via the FIB, it is replicated for
each next hop and sent via each next hop's \fname{send()} function.

If the message is a Content Object, it is matched against the PIT.  If a hit is found, the message
is replicated for each previous hop and sent to that connection's \fname{send()} function.

Finally, if the message is a Control packet, it is sent to the Configuration module.
If the message is not any of a Content Object, Interest, or Control it is dropped.

Metis currently does not implement the \emph{InterestReturn} message.

\subsubsection{PIT Table}\label{sec:pit}
Metis includes one PIT implementation, \fname{MetisStandardPIT} which implements the
\fname{MetisPIT} interface, shown in Figure~\ref{fig:MetisPIT}.  When the Message Processor
receives an Interest, it calls \fname{receiveInterest()} and the PIT table returns a PIT Verdict
indicating if the Interest should is aggregated or should be forwarded.  When the Message
Processor receives a Content Object, it calls \fname{satisyInterest()} and gets back a
set of Connection IDs to forward the Content Object to.  The list could be empty if no
match is found.

\begin{figure}
\begin{lstlisting}[language=C]
struct metis_pit {
    void (*release)(MetisPIT **pitPtr);
    MetisPITVerdict (*receiveInterest)(MetisPIT *pit, MetisMessage *interestMessage);
    MetisNumberSet * (*satisfyInterest)(MetisPIT *pit, const MetisMessage *objectMessage);
    void (*removeInterest)(MetisPIT *pit, const MetisMessage *interestMessage);
    MetisPitEntry * (*getPitEntry)(const MetisPIT *pit, const MetisMessage *interestMessage);
    void *closure;
};
\end{lstlisting}
\caption{MetisPIT Interface}
\label{fig:MetisPIT}
\end{figure}

\subsubsection{FIB Table}\label{sec:fib}
TBD

\subsubsection{ContentStore}\label{sec:contentstore}
A ContentStore implements the \fname{MetisContentStoreInterface}, shown in Figure~\ref{fig:ContentStore}.
When the Message Processor receives a Content Object that it wishes to cache, it calls \fname{putContent()},
which may evict an older item.  The eviction policy is up to the content store implementation.
When the Message Processor receives an Interest that is not already in the PIT, it tries to satisfy
it by calling \fname{matchInterest}.  If a match is found, it returns the MetisMessage of the corresponding Content Object.

\begin{figure}
\begin{lstlisting}[language=C]
struct metis_contentstore_interface {
    bool (*putContent)(MetisContentStoreInterface *storeImpl, MetisMessage *content, uint64_t currentTimeTicks);
    bool (*removeContent)(MetisContentStoreInterface *storeImpl, MetisMessage *content);
    MetisMessage * (*matchInterest)(MetisContentStoreInterface*storeImpl, MetisMessage *interest);
    size_t (*getObjectCapacity)(MetisContentStoreInterface *storeImpl);
    size_t (*getObjectCount)(MetisContentStoreInterface *storeImpl);
    void (*log)(MetisContentStoreInterface *storeImpl);
    MetisContentStoreInterface *(*acquire)(const MetisContentStoreInterface *storeImpl);
    void (*release)(MetisContentStoreInterface **storeImpl);
    void *_privateData;
};\end{lstlisting}
\caption{ContentStore Interface}
\label{fig:ContentStore}
\end{figure}

\subsection{Connection Manager}
The Connection Manager is a \fname{MetisMissive} listener.  When it receives connection event messages, it
forwards them to applications that have registered (TBD) and cleans up the Connection Table for
connections that have gone away.

For example, when the Connection Manager receives a CLOSED signal for a connection, it will
remove that connection from the connection table and remove it as a next hop from all routes.

The connection manager queues received Missives and processes them in a later Dispatcher
scheduling time.  This avoid conflict with other Missive receivers.

\section{Programming Tasks}
This section describes how to modify certain components of Metis to evaluate different technologies
or change the behavior.  The modular pieces are the PIT, FIB, Content Store, and protocol Listeners and
IO Connections.

\subsection{Replacing the PIT Table}
A PIT table must implement the \fname{MetisPIT} interface.  The included \fname{MetisStandardPIT} implements
the CCNx 1.0 specification for the PIT table.

To replace the standard PIT with a customized PIT, change the call to \fname{metisStandardPIT\_Create()}
in \fname{metisMessageProcessor\_Create()} to the new constructor.  There should be no additional changes.

There is currently no means to choose a PIT table implementation by configuration.

\subsection{Replacing the Content Store}
A Content Store must implement the \fname{MetisContentStoreInterface} interface.  In the
\fname{metisMessageProcessor\_Create()} function, simply replace the call to 
\fname{metisLRUContentStore\_Create()} with your own Content Store implementation.

Metis allows the size of the Content Store to be set via configuration.  This results in a call
to \fname{metisMessageProcessor\_SetContentStoreSize()}.  You should edit this function to
use whatever means you implement for a replacement Content Store.  The LRU ContentStore
simply releases itself and creates a new one, which does result in losing all cached content.

There is currently no means to choose a Content Store implementation by configuration.

\subsection{Adding a new I/O Protocol}
An I/O Protocol has four pieces: the ProtocolListener, the ProtocolConnection, the ProtocolTunnel,
and the ProtocolConfiguration.  The first three pieces live in the ``io'' directory and the configuration
piece lives in the ``config'' directory.

For purposes of explanation, lets use SCTP as an example new protocol.  The new modules to add
to Metis would be \fname{SCTPListener}, \fname{SCTPConnection}, \fname{SCTPTunnel}, and
add configuration options to \fname{metisControl\_AddListener}, \fname{metisControl\_RemoveListener},
\fname{metisControl\_AddConection}, \fname{metisControl\_RemoveConection}, 
\fname{metis\_Configuration}, and \fname{metis\_ConfigurationListeners}.

The \fname{MetisConfiguation} components will be refactored to allow a more modular approach
to adding protocols.

\subsubsection{The I/O pieces}
The protocol listener, in our example \fname{SCTPListnener}, sets up the server socket for the protocol.
The listener would function much like the UDP listener, using \fname{bind(), listen(), and recvmsg()}
with a \texttt{SOCK\_SEQPACKET} socket type.  It would accept packets and determine if it matched
an existing connection.  If not, it would create a connection and \fname{SCTPConnection} object to
put in the Connection Table.

Because one would want to send a reply packet from the server socket address, the \fname{SCTPConnection}
would use the same socket as the \fname{SCTPListener}.

The \fname{SCTPTunnel} module is used by the configuration system to create an out-bound connection
to a remote system.  Like the \fname{UDPTunnel}, its main job is to lookup the appropriate \fname{SCTPListener}
-- so it can borrow the socket -- and then create a \fname{SCTPConnection} and put it in the connection table.

An alternate approach would be use SCTP in one-to-one mode, in which case it would follow
the \fname{TCPListener, TCPConnection} model.

\subsubsection{The Configuration pieces}
The configuration process requires updates to each of these sections to enable configuration via \control and a configuration file.
\begin{description}
\item[\fname{metisControl\_AddListener}] {Define the ``ADD LISTENER'' command syntax for the listener.}
\item[\fname{metisControl\_RemoveListener}] {Define the ``REMOVE LISTENER'' command syntax for the listener.}
\item[\fname{metisControl\_AddConection}] {Define the ``ADD CONNECTION'' command syntax.  For IP based protocols,
it will likely fall in to the \fname{\_metisControlAddConnection\_ParseIPCommandLine} format and use the
\fname{\_metisControlAddConnection\_IpHelp} help display.}
\item[\fname{metisControl\_RemoveConection}] {Define the ``REMOVE CONNECTION'' command syntax.}
\end{description}

The result of these \fname{metisControl\_X} functions is a CPI control object that can be sent down
the protocol stack and encoded to Metis for configuration.  These code modules create
a \fname{CCNxMetaMessage} and pass it to \fname{ccnxControlState\_WriteRead()}.  

The \fname{ccnxControlState\_WriteRead()} function is program specific.  For a program like \control,
it will result in the \fname{CCNxMetaMesage} being sent down the protocols stack to in to Metis via
a network channel.  For parsing the configuration file within Metis, it will result in the message
being handed off directly to \fname{MetisConfiguration}.

\begin{description}
\item[\fname{metisConfiguration\_ProcessCreateTunnel}] Add a handler to \fname{SCTPTunnel\_Create()}.
\item[\fname{metisConfiguration\_ProcessRemoveTunnel}] Add a handler to move the connection to CLOSED state.
\item[\fname{metisConfigurationListeners\_Add}] Add a handler to \fname{SCTPListener\_Create()}..
\item[\fname{metisConfigurationListeners\_Remove}] Add a handler to close all connections using
the listener and remove the Listener.
\end{description}

\bibliographystyle{unsrt}
\bibliography{sample}


\end{document}